\documentclass[12pt]{article}
\usepackage{psfig}
\usepackage{aaspp4}

\def\lsim{\mathrel{\hbox{\rlap{\lower.55ex \hbox {$\sim$}}\kern-.0em
\raise.4ex \hbox{$<$}}}} 
\def\gsim{\mathrel{\hbox{\rlap{\lower.55ex \hbox {$\sim$}}\kern-.0em
\raise.4ex \hbox{$>$}}}} 
\newcommand{\gpm}[3]{$#1^{+#2}_{-#3}$}

\def\grb{GRB\thinspace{991208}}

\begin{document}

\title{The Bright Gamma-Ray Burst 991208 -- Tight Constraints on
Afterglow Models from Observations of the Early-Time Radio Evolution}

\author{
T. J. Galama\altaffilmark{1},
M. Bremer\altaffilmark{2},
F. Bertoldi\altaffilmark{3},
K.M. Menten\altaffilmark{3},
U. Lisenfeld\altaffilmark{4},
D. S. Shepherd\altaffilmark{5},
B. Mason\altaffilmark{6},
F. Walter\altaffilmark{6},
G. G. Pooley\altaffilmark{7},
D. A. Frail\altaffilmark{5},      
R. Sari\altaffilmark{8},        
S. R. Kulkarni\altaffilmark{1},       
E. Berger\altaffilmark{1},     
J.S. Bloom\altaffilmark{1},
A. J. Castro-Tirado\altaffilmark{9},
J. Granot\altaffilmark{10}}
\vspace{-3mm}

\altaffiltext{1}{Division of Physics, Mathematics and Astronomy,
 California Institute of Technology, MS~105-24, Pasadena, CA 91125}

\altaffiltext{2}{Institut de Radio Astronomie Millim\'{e}trique, 300
rue de la Piscine, F--38406 Saint-Martin d'H\`{e}res, France}

\altaffiltext{3}{Max-Planck-Institut f\"ur Radioastronomie, Auf dem
H\"ugel 69, D--53121 Bonn, Germany}

\altaffiltext{4}{Instituto de Radioastronomia Milimetrica, Avenida
    Pastora 7, Nucleo Central, E-18012 Granada, Spain}

\altaffiltext{5}{National Radio Astronomy Observatory, P.O. Box 0,
Socorro, NM 87801}

\altaffiltext{6}{California Institute of Technology,
Owens Valley Radio Observatory 105-24, Pasadena, CA 91125}    

\altaffiltext{7}{Mullard Radio Astronomy Observatory,
Cavendish Laboratory, University of Cambridge, Madingley Road,
Cambridge CB3 0HE, UK}

\altaffiltext{8}{California Institute of Technology,
 Theoretical Astrophysics  103-33, Pasadena, CA 91125}
                                                               
\altaffiltext{9}{LAEFF-INTA, Villafranca del Castillo, PO Box 50.727,
E--28080 Madrid, Spain }

\altaffiltext{10}{AC Racah Institute, Hebrew University, Jerusalem
91904, Israel}

\begin{abstract}
The millimeter wavelength emission from \grb\ is the second brightest
ever detected, yielding a unique data set. We present here well-sampled
spectra and light curves over more than two decades in frequency for a
two-week period.  This data set has allowed us for the first time to
trace the evolution of the characteristic synchrotron self-absorption
frequency $\nu_{\rm a}$ and peak frequency $\nu_{\rm m}$, and the peak
flux density F$_{\rm m}$: we obtain $\nu_{\rm a} \propto t^{-0.15 \pm
0.12}$, $\nu_{\rm m} \propto t^{-1.7 \pm 0.4}$, and F$_{\rm m} \propto
t^{-0.47 \pm 0.11}$.  From the radio data we find that models of
homogeneous or wind-generated ambient media with a spherically
symmetric outflow can be ruled out. A model in which the relativistic
outflow is collimated (a jet) can account for the observed evolution
of the synchrotron parameters, the rapid decay at optical wavelengths,
and the observed radio to optical spectral flux distributions that we
present here, provided that the jet transition has not been fully
completed in the first two weeks after the event. These observations
provide additional evidence that rapidly decaying optical/X-ray
afterglows are due to jets and that such transitions either develop
very slowly or perhaps never reach the predicted asymptotic decay
F$(t) \propto t^{-p}$.
\end{abstract} 



\section{Introduction}

GRB\,991208 was detected with the Interplanetary Network (IPN) by the
{\em Ulysses}, {\em WIND} and {\em NEAR} spacecraft on December 8,
1999, at 04:36:52 UT.  The gamma-ray properties, the localization and
the subsequent discovery of the radio counterpart to GRB\,991208 are
presented by Hurley {\it et al.}  (2000)\nocite{hcm+00}.



The IPN can typically localize events within one or two days.
Therefore, important spectral transitions, such as the passage of the
synchrotron peak and cooling frequencies (\cite{rkf+98,gwb+98}), that
occur at optical and infrared wavelengths a few hours after the event
are missed by optical and infrared follow-up observations on IPN
bursts. Fortunately, the radio range allows similar studies on time
scales from days to weeks.

Here we present an extensive set of radio observations monitoring
GRB\,991208 between 1.4 and 250 GHz during the first two weeks after
the event. It is the second brightest afterglow detected at millimeter
wavelengths to date (\cite{she+99}). The brightness of the afterglow
allows us to trace the evolution of the characteristic synchrotron
peak frequency, the self-absorption frequency, and the peak flux
density.  The observations and results are presented in \S
\ref{sec:results}, and compared with appropriate
models in \S \ref{sec:dis}.

\section{Observations and results}
\label{sec:results}

Observations were made from 1.43 to 250 GHz, at a number of
facilities: the Max-Planck Millimeter Bolometer (MAMBO; \cite{kgg+98})
at the IRAM 30-m telescope on Pico Veleta (Spain; \cite{bhmj87}), the
IRAM Plateau de Bure Interferometer (PdBI) in the French Alps
(\cite{gdd+92}), the Owens Valley Radio Observatory (OVRO)
millimeter-wave array, the OVRO 40-meter telescope, the Ryle Telescope
at Cambridge (UK), and the NRAO Very Large Array
(VLA)\footnotemark\footnotetext{The NRAO is a facility of the National
Science Foundation operated under cooperative agreement by Associated
Universities, Inc.}. We have detailed our observing and calibration
strategy in Kulkarni {\it et al.} (1999)\nocite{kfs+99}, Galama {\it
et al.} (1999)\nocite{gbw+99}, Frail {\it et al.}
(2000)\nocite{fbg+00}, and Berger {\it et al.} (2000)\nocite{bsf+00}.
A log of the observations is provided in Table \ref{tab:obs}.

\begin{figure}[b!] 
\centerline{\psfig{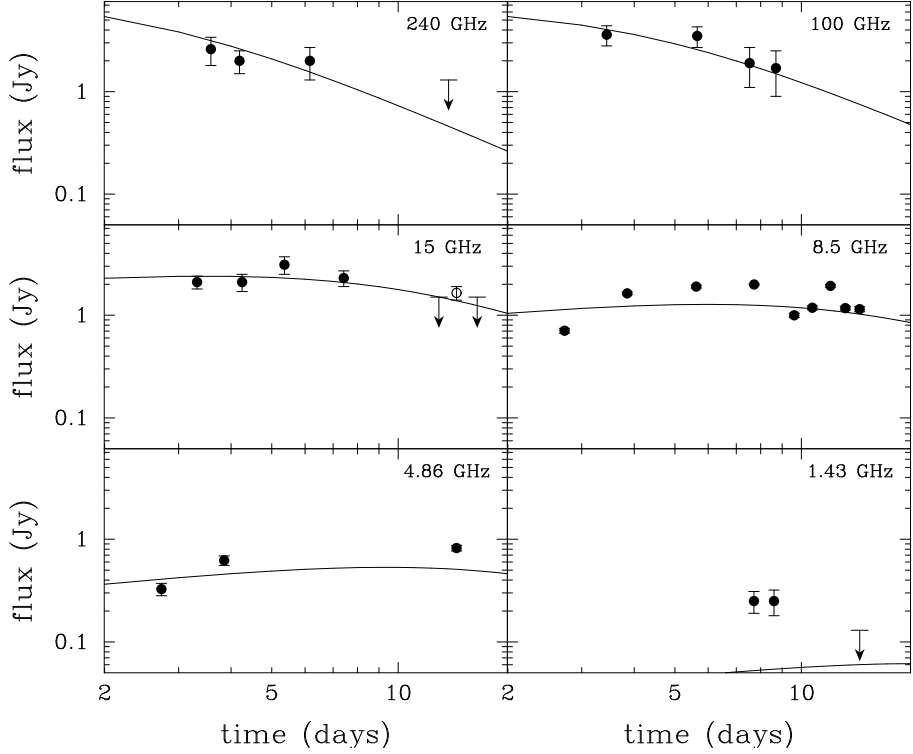}}
\caption{Radio, mm and sub-mm light curves of GRB\,981208 from 1.4 to
250 GHz.  These data are compiled in Table \ref{tab:obs} and are taken
at various telescopes which include the IRAM 30-m (250 GHz), the OVRO
Millimeter Array (100 GHz), the Ryle Telescope (15 GHz), and the VLA
(8.5, 4.9 and 1.4 GHz). At 15 GHz there is one additional point on the
plot (open circle) that was taken with the VLA. All upper limits are
plotted as the peak flux density at the location of the afterglow plus
two times the rms noise in the image. Frequencies below 10 GHz show
erratic flux density variations due to ISS.  Shown are the lightcurves
that result from assuming an evolution of $\nu_{\rm a}$, $\nu_{\rm
m}$, and F$_{\rm m}$ as derived in \S \ref{sec:results},
i.e. $\nu_{\rm a} \propto t^{-0.15 \pm 0.12}$, $\nu_{\rm m} \propto
t^{-1.7 \pm 0.4}$, and F$_{\rm m} \propto t^{-0.47 \pm 0.11}$ and
using the Granot {\it et al.} (1999a, 1999b) formulation.}
\label{fig:obs}
\end{figure}

The millimeter emission from \grb\ is one of the brightest ever
detected (\cite{bbg+98,stv+98,sfkm98}) enabling us to obtain a unique
data set. Rather than single-epoch snapshot spectra (\cite{gwb+98}), we
have well-sampled spectra and light curves over more than two decades
in frequency for a two-week period.  GRB\,991208 is therefore
well-suited to a study of the broad-band evolution of the radio
afterglow in the first two weeks after the burst. We display light
curves for a subset of these data in Fig. \ref{fig:obs}. The data from
Table \ref{tab:obs} at 22 GHz, 30 GHz and 86 GHz are too sparse to
plot in this form.

Several trends are apparent. At the highest frequencies (250 GHz, 100
GHz and 15 GHz) the flux densities are either declining or constant
during the first week, but thereafter fade below detectability. At
frequencies below 10 GHz we see erratic flux density
variations due to interstellar scintillation (ISS; see
e.g. \cite{goo97}; details on the
ISS properties of GRB\,991208 can be found in Galama {\it et al.}
2000)\nocite{gal+00}. Therefore, below 10 GHz it is harder to discern
a pattern. The overall trend in Fig. \ref{fig:obs} is for the peak
flux density to decline with decreasing frequency, while the
time-to-maximum increases.

\begin{figure}[b!] 
\centerline{\psfig{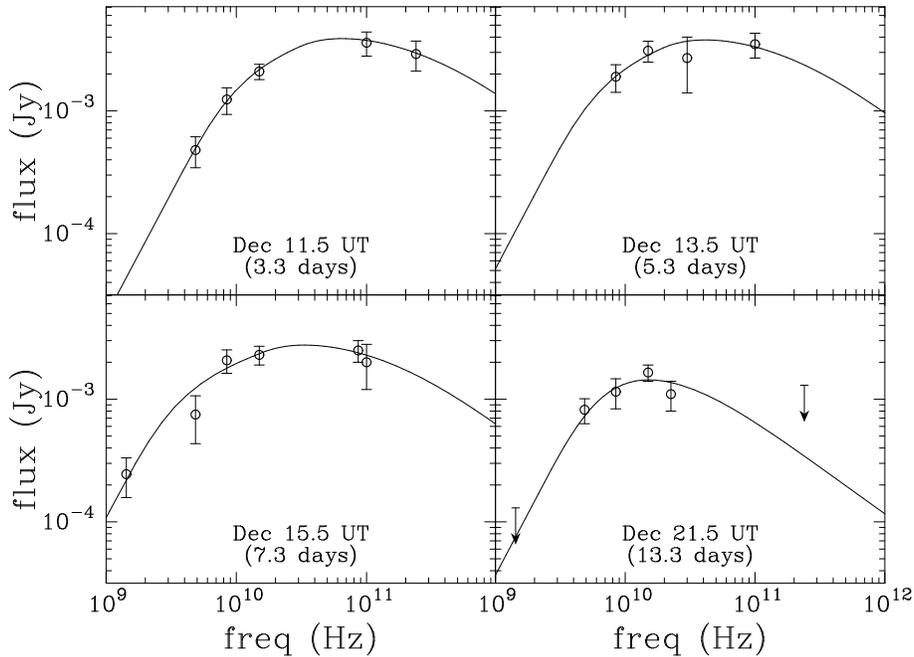}}
\caption{The radio spectral flux distribution of GRB\,991208
from 1.4 to 250 GHz at several epochs. Shown are fits of the
synchrotron spectra to the radio data from a
relativistic blast wave as specified by Granot {\it et al.}~(1999a,
1999b) for $p = 2.52$. }
\label{fig:spec}
\end{figure}

An alternate way to view the data in Table \ref{tab:obs} is to
construct instantaneous spectral flux distributions at several epochs,
which were chosen to obtain a maximum number of observations
sufficiently close in time. All values were brought to the same epoch
by applying a correction that was determined by fitting cubic splines
to each of the light curves in Fig.  \ref{fig:obs}. We have added in
quadrature an additional error of 25~\% of the synchrotron spectral
model flux (see \S \ref{sec:specfits}) to the fluxes below 10 GHz to
reflect the uncertainty due to ISS (see \cite{gal+00}). The resulting
spectral flux distributions are presented in Fig. \ref{fig:spec}. The
spectra for all four epochs show the same overall morphology - flat
from 10 GHz to 250 GHz, and dropping precipitously below 10 GHz.

In addition, we reconstruct the radio to optical spectral flux
distribution at the epoch of December 15.5 UT by including K, R and
I-band detections (\cite{bdk+99,mpp+99b,smp+00}). All values were
brought to the same epoch by applying a correction using the slope of
the fitted optical light curve (\cite{smp+00}). We corrected for
Galactic foreground reddening (A$_R = 0.043$, A$_I = 0.031$, and A$_K
= 0.006$; as inferred from the dust maps of Schlegel, Finkbeiner and
Davis 1998\nocite{sfd98}\footnote{see
http://astro.berkeley.edu/davis/dust/index.html}). The result is
presented in Fig. \ref{fig:radiotoopt}. 

\subsection{Spectral fits \label{sec:specfits}}

The cm/mm/optical to X-ray afterglow emission is believed to arise
from the forward shock of a relativistic blast wave that propagates
into the circumburst medium (see \cite{pir99,vkw00} for reviews). 
The synchrotron afterglow spectrum has four distinct regions: $F_{\nu}
\propto \nu^{2}$ below the synchrotron self-absorption frequency
$\nu_{\rm a}$ ($\nu < \nu_{\rm a}$); $F_{\nu} \propto \nu^{1/3}$ up to
the synchrotron peak at $\nu_{\rm m}$ ($\nu_{\rm a} < \nu < \nu_{\rm
m}$); $F_{\nu} \propto \nu^{-(p-1)/2}$ above the peak up to the
cooling frequency $\nu_{\rm c}$ ($\nu_{\rm m} < \nu < \nu_{\rm c}$);
and $F_{\nu} \propto \nu^{-p/2}$ above cooling ($\nu > \nu_{\rm c}$;
see e.g. \cite{spn98}; we have assumed slow cooling, i.e. $\nu_{\rm m}
< \nu_{\rm c}$). Here, $p$ is the power-law index of the
electron energy distribution. 

The spectral flux distribution of Dec 15.5 UT
(Fig. \ref{fig:radiotoopt}) peaks at $\nu_{\rm m} \sim$ 40
GHz. 
A fit to the radio data is provided by a synchrotron spectrum from a
relativistic blast wave as specified by Granot, Piran and
Sari~(1999a\nocite{gps99a}; see their Fig. 10 for the equipartition
field model). We scale their dimensionless functional form by peak
frequency $\nu_{\rm m}$ and peak flux density $F_{\rm m}$ to derive a
function $g(\nu)$ with asymptotic behavior of $\nu^{1/3}$ ($\nu <<
\nu_{\rm m}$) and $\nu^{-(p-1)/2}$ ($\nu >> \nu_{\rm m}$). We account
for synchrotron self-absorption at $\nu_{\rm a}$ by multiplying
$g(\nu)$ by $F_\nu=[1-{\rm exp}(-\tau)]/\tau$, where
$\tau=(\nu/\nu_{\rm a})^{-5/3}$ (\cite{gps99b}).

First we fit the radio to optical spectral flux distribution of
December 15.5 UT and find: $p = 2.524 \pm 0.016$, $\nu_{\rm a} = $
\gpm{3.6}{1.1}{0.8} GHz, $\nu_{\rm m} = 35  \pm 4$ GHz, and F$_{\rm
m} = 2.85 \pm 0.15$ mJy ($\chi_{\rm r}^2 = 1.6$; 6 degrees of freedom;
d.o.f.). The spectrum is well described by synchrotron emission from a
power-law electron distribution with index $p = 2.52$, self-absorbed at
low frequencies; the fit is shown in Fig. \ref{fig:radiotoopt}.

Next, we fix $p = 2.52$ and fit all the epochs of the radio spectral
flux distributions (the fits are shown in Fig. \ref{fig:spec}): the
derived values of $\nu_{\rm a}$, $\nu_{\rm m}$ and $F_{\rm m}$ and
their uncertainties are given in Table \ref{tab:fits}. These are
plotted in Fig. \ref{fig:fit}, where a clear temporal evolution of
$\nu_{\rm m}$ and $F_{\rm m}$ is apparent. We note that assuming $p =
2.2$ (see \S \ref{sec:dis}) does not change the results of the fits
(i.e., when no optical data is included the fits are rather
insensitive to the assumed value of $p$). To characterize this
evolution, we made power-law least squares fits to the data. The
results are given in Table \ref{tab:fitsres}.

\section{Discussion}\label{sec:dis}

The brightness of the radio counterpart to GRB\,991208
has allowed for the first time to trace the evolution of the
characteristic synchrotron parameters $\nu_{\rm a}$, $\nu_{\rm m}$ and
F$_{\rm m}$. We may now compare this result directly with expectations
from relativistic blast wave models. 

Currently popular scenarios for the origin of GRBs are: a compact
object merger (the neutron star-neutron star, \cite{eic+89}, and
neutron star-black hole merger models, \cite{ls74,npp92}), and the
core collapse of a very massive star (`failed' supernova or hypernova
\cite{woo93,pac98b}).  In principle afterglow observations can
distinguish a compact object merger, which is expected to occur in a
constant density interstellar medium (constant density ISM model; see
\cite{spn98} and \cite{wg99} for details), from a hypernova, where the
circumburst environment will have been influenced by the strong wind
of the massive progenitor star (wind model; see \cite{cl99,cl99b}).
Below we compare each of the afterglow models separately with the
observations. The expected scalings (we assume adiabatic expansion of
the remnant) for each of the models, including the case where the
outflow is collimated in a jet, are also provided in Table
\ref{tab:fitsres} for comparison with the evolution derived from the
spectral fits.

\begin{figure}[b!] 
\centerline{\psfig{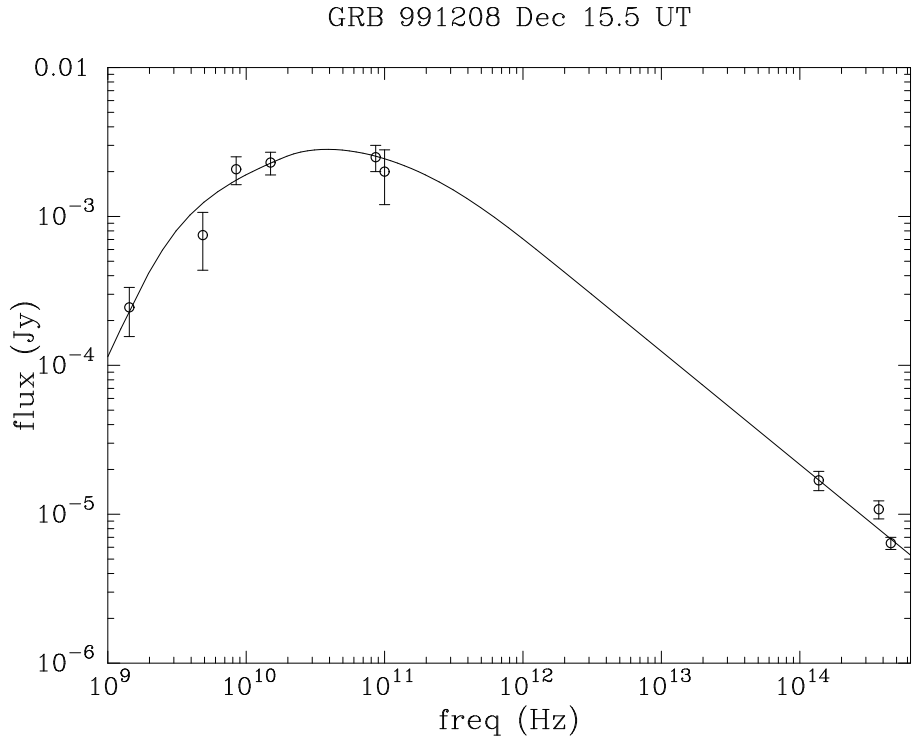}}
\caption{The radio to optical spectral flux distribution of
GRB\,991208 at Dec 15.5 UT (7.3 days after the event). Shown is a fit
of the synchrotron spectrum to the radio and infrared/optical data
from a relativistic blast wave as specified by Granot {\it et
al.}~(1999a, 1999b) for $p = 2.52$.}
\label{fig:radiotoopt}
\end{figure}

The observed decay of the peak-flux density F$_{\rm m}$ is
inconsistent with the constant density ISM model (see Table
\ref{tab:fitsres}), which predicts a constant peak flux.  Also, the
optical afterglow of GRB\,991208 (R band) displays a rapid decay:
$\alpha = -2.2 \pm 0.1$, $F_{\nu} \propto t^{\alpha}$ (\cite{smp+00};
see also \cite{cas+00}). A constant density ISM model would require $p
\geq 3.6$ in order to account for the rapid decay observed at optical
wavelengths. For $p \geq 3.6$ the synchrotron spectrum above the peak
should be much steeper than observed ($F_{\nu} \propto \nu^{-(p -
1)/2}$). We note that optical extinction local to the host would cause
the observed spectrum to be even shallower. The constant density ISM
model can thus be ruled out on the basis of the spectral flux
distribution on 1999 Dec. 15.5 UT, which shows a flat spectrum
corresponding to $p = 2.52$ (see Fig. \ref{fig:radiotoopt}).

Comparing GRB\,991208 with the wind model (see Table
\ref{tab:fitsres}) we find that the evolution of the peak-flux density
F$_{\rm m}$ and the peak frequency $\nu_{\rm m}$ are as expected, but
we find no evidence for a decrease with time of the self-absorption
frequency $\nu_{\rm a}$. And, similar to the case of the constant
density ISM model, a wind model would require $p \geq 3.3$ in order to
account for the rapid decay observed at optical wavelengths. We
therefore also reject the wind model.

Finally, the outflow may be collimated into jets (jet model; see for
details \cite{rho99,sph99,pm99}). Once sideways expansion of the jet
dominates the dynamics of the relativistic-blast wave, a jet plus
constant ambient density medium (ISM + jet) or a jet plus wind
stratified ambient medium (wind + jet) will have very similar
properties. In this stage, the morphology of the ambient medium will
be difficult to constrain.

The observed evolution of $\nu_{\rm a}$, and $\nu_{\rm m}$ are as
expected in the jet model, but the drop in peak-flux density F$_{\rm
m}$ is not as rapid as predicted (see Table \ref{tab:fitsres}).  To
obtain the predicted decay rate F$_{\rm m} \propto t^{-1}$ we have to
raise $\chi^2$ by 21.0 (corresponding to $> 4 \sigma$); the observed
decay is thus significantly different from the predicted decay.  And,
in the case of a jet, at late times, when the evolution is dominated
by the spreading of the jet, the model predicts $\alpha = -p$, and the
R-band measurement thus corresponds to $p = 2.2 \pm 0.1$. For $p =
2.2$ the synchrotron spectrum above the peak should be shallower than
observed. But, we find that $p = 2.52$ gives a satisfactory
description of the Dec 15.5 UT data (Fig. \ref{fig:radiotoopt}). For
example assuming $p = 2.2$ provides a worse fit ($\chi^{2}_{\rm r}$ =
3.4 as opposed to 1.6 for $p = 2.52$; Dec. 15.5 UT radio to optical
spectral flux distribution; Fig. \ref{fig:radiotoopt}).

A $p = 2.2$ jet model could only be saved if one assumes a cooling
break between the mm and optical passbands, around $\nu_{\rm c} =
5\times10^{13}$ Hz (the optical decay would still satisfy $\alpha =
-p$, i.e. $p = 2.2 \pm 0.1$). However, in that case the
optical/infrared synchrotron spectrum F$_{\nu} \propto \nu^{-p/2} =
\nu^{-1.1}$, i.e. significantly steeper than observed ($\beta = -0.75
\pm 0.03$, F$_\nu \propto \nu^\beta$; \cite{smp+00}). We therefore
reject the possibility that $p = 2.2$.

In the jet model, the shallow decay of the peak-flux density F$_{\rm
m}$ and the apparent discrepancy in the value of $p$ derived from the
spectrum and from the light curve, can be understood if during the
first two weeks the GRB remnant was still undergoing the transition
from a quasi-spherical (either constant density ISM or a wind),
i.e. F$(t) \propto t^{-(3p-3)/4} \sim t^{-1.1}$ (constant density
ISM), or F$(t) \propto t^{-(3p-1)/4} \sim t^{-1.6}$ (wind) to a jet
evolution, F$(t) \propto t^{-p} \sim t^{-2.5}$. As discussed by Kumar
and Panaitescu (2000) this transition takes place over at least an
order of magnitude, but more likely several orders of magnitudes, in
observer time. Since typically we expect the jet transition to take
place around one to several days (\cite{sph99}) it is natural to
expect that the transition was still ongoing; the evolution of the
synchrotron parameters will then be somewhere between fully spherical
at early times (either constant density ISM or a wind) and fully
dominated by a jet later. This is consistent with the data, would
explain the shallow decay of the peak flux density, and would also
explain why the optical decay rate exponent $\alpha = -2.2$
(\cite{smp+00}) is not as steep as for a fully developed jet
transition $\alpha = -p = -2.52$. Determination of $p$ from light
curves requires observations after the transition has been completed
and from the optical data we cannot establish whether this is the
case. Also the light curve is not necessarily expected to approach the
asymptotic limit $\alpha = -p$.  The value of $p$ derived from the
optical light curve is thus rather uncertain and the value of $p =
2.52$ that we have inferred from the spectral flux distribution is the
more robust number, since $p$ so derived is independent of the
hydrodynamical evolution of the blastwave.

A similarly slowly developing jet transition was observed for
GRB\,990123 (\cite{kdo+99,ftm+99,crc+99}), where after the jet break
(at $t_{\rm J} \sim 2$ days) the optical decay rate exponent $\alpha
\sim -1.8$; whereas a steep $\alpha = -p = -2.4$ would have been
expected for a fully developed jet transition. Also for GRB\,991216 a
slow transition is seen, but due to host contamination, the late-time
decay rate exponent can only be constrained to lie between $-2.1$ and
$-1.5$ (\cite{hum+00}).  On the other hand GRB\,990510 and GRB\,000301C
appear to be examples of relatively fast and fully developed jet
transitions (\cite{hbf+99,bsf+00}; but see \cite{kp00}).

In conclusion, the jet model can account for the observed evolution of
the synchrotron parameters of GRB\,991208, the rapid decay at optical
wavelengths and for the observed radio to optical spectral flux
distributions, provided that the jet transition has not been fully
completed in the first two weeks after the event. These observations
provide additional evidence that rapidly decaying afterglows are due
to jets (\cite{sph99}) and that at least in some afterglows these
transitions are not fully developed.

\begin{figure}[b!] 
\centerline{\psfig{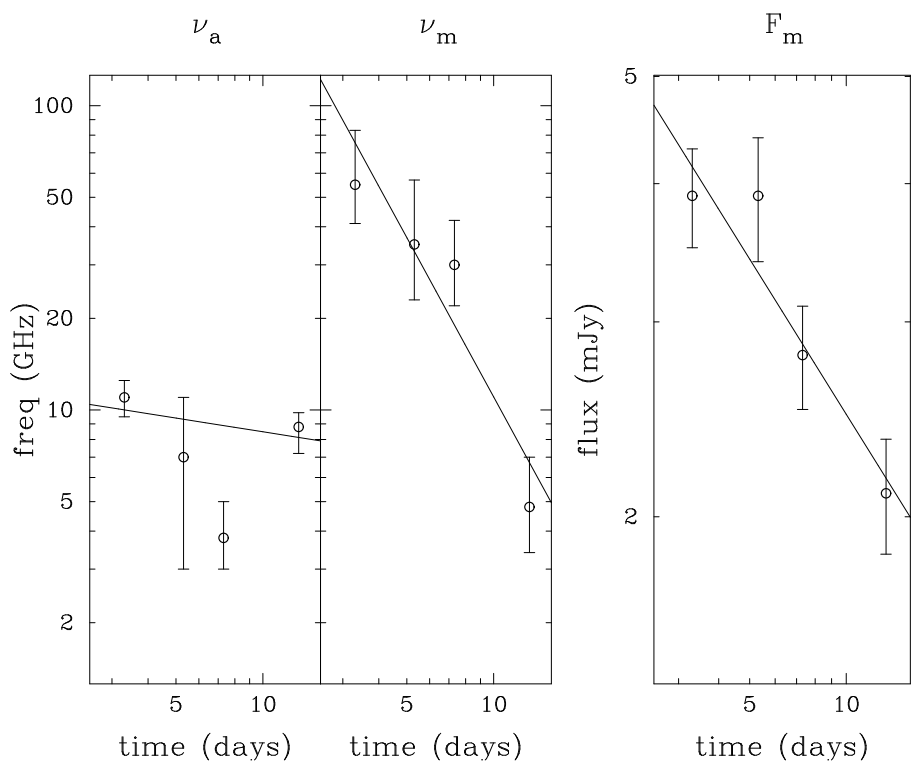}}
\caption{Evolution of the the synchrotron self-absorption frequency
$\nu_{\rm a}$, the synchrotron peak frequency $\nu_{\rm m}$ and the
peak flux density F$_{\rm m}$ as derived from broad band radio (from
1.4 to 250 GHz) spectral fits of GRB\,981208 at several epochs. Shown
are power-law fits to the data: $\nu_{\rm a} \propto t^{-0.15 \pm
0.12}$, $\nu_{\rm m} \propto t^{-1.7 \pm 0.4}$, and F$_{\rm m} \propto
t^{-0.47 \pm 0.11}$.  The fitted spectra are show in \ref{fig:spec},
and the respective fit parameters can be found in Table
\ref{tab:fits}.}
\label{fig:fit}
\end{figure}

Around 10 days after the event $\nu_{\rm m}$ passed $\nu_{\rm a}$
($\nu_{\rm m} < \nu < \nu_{\rm a}$; see Fig. \ref{fig:fit}) at $\sim$
8 GHz; from then on $\nu_{\rm a}$ is expected to evolve differently:
$\nu_{\rm a} \propto t^{-2(1 + p)/(4 + p)} = t^{-14/13}$ (for $p =
2.5$); we have assumed a jet-model evolution.  Also after $\nu_{\rm
m}$ has passed 8.46 ($\sim$ 12 days) and 4.86 GHz ($\sim$ 17 days),
the flux densities at these frequencies should decay rapidly F$_{\nu}
\propto t^{-2.2-2.5}$. Continued monitoring with the VLA
(\cite{gal+00}) will allow tests of these expectations.


\acknowledgements Research at the Owens Valley Radio Observatory is
supported by the National Science Foundation through NSF grant number
AST 96-13717.  We would like to thank S. Jogee for preparing the first
observations of this burst at the Owens Valley Observatory and A.
Sargent for generously allocating the time necessary to track the
millimeter evolution. We thank A. Diercks for useful discussions.



\footnotesize
\begin{deluxetable}{llll}
\tabcolsep0in\footnotesize \tablewidth{\hsize} \tablecaption{Radio
Observations of \grb\label{tab:obs}} \tablehead { \colhead {Date} &
\colhead {Freq.} & \colhead {Flux} & \colhead {Remarks} \\ \colhead
{1999 Dec. (UT)} & \colhead {(GHz)} & \colhead {($\mu$Jy)} & \colhead
{ } } \startdata 11.77 & 250 & 2600 $\pm$ 800 & MAMBO, $\tau_{atm}$ =
0.16\\ 12.38 & 250 & 2000 $\pm$ 500 & MAMBO, $\tau_{atm}$ = 0.22 \\
14.35 & 250 & 2000 $\pm$ 700 & MAMBO, $\tau_{atm}$ = 0.54-0.77\\ 21.38
& 250 & $\phantom{0}$100 $\pm$ 600 & MAMBO, $\tau_{atm}$ = 0.16\\
11.63 & 100 & 3600 $\pm$ 800 & OVRO MA\\ 13.84 & 100 & 3500 $\pm$ 800
& OVRO MA\\ 15.72 & 100 & 1900 $\pm$ 800 & OVRO MA\\ 16.89 & 100 &
1700 $\pm$ 800 & OVRO MA\\ 15.62 & 86.24& 2500 $\pm$ 500 & PdBI\\
13.63 & 30.0 & 2700 $\pm$ 1300 & OVRO 40-m \\ 14.62 & 30.0 & 1700
$\pm$ 1000 & OVRO 40-m \\ 14.84 & 30.0 & 3600 $\pm$ 1200 & OVRO 40-m
\\ 11.51 & 15.0 & 2100 $\pm$ 300 & Ryle \\ 12.44 & 15.0 & 2100 $\pm$
400 & Ryle \\ 13.55 & 15.0 & 3100 $\pm$ 600 & Ryle \\ 15.61 & 15.0 &
2300 $\pm$ 400 & Ryle \\ 20.68 & 15.0 & $\phantom{0}$300 $\pm$ 600 &
Ryle \\ 23.60 & 15.0 & $\phantom{0}$500 $\pm$ 500 & Ryle \\ 21.96 &
22.49 & 1100 $\pm$ 300 & VLA \\ 21.96 & 14.97 & 1650 $\pm$ 250 & VLA
\\ 10.92 & 8.46 & $\phantom{0}$707 $\pm$ 39 & VLA \\ 12.04 & 8.46 &
1630 $\pm$ 51 & VLA \\ 13.81 & 8.46 & 1900 $\pm$ 66 & VLA \\ 15.91 &
8.46 & 1990 $\pm$ 33 & VLA \\ 17.81 & 8.46 & $\phantom{0}$999 $\pm$ 50
& VLA \\ 18.81 & 8.46 & 1182 $\pm$ 52 & VLA \\ 19.92 & 8.46 & 1930
$\pm$ 53 & VLA \\ 20.92 & 8.46 & 1173 $\pm$ 49 & VLA \\ 21.96 & 8.46 &
1146 $\pm$ 66 & VLA \\ 10.92 & 4.86 & $\phantom{0}$327 $\pm$ 45 & VLA
\\ 12.04 & 4.86 & $\phantom{0}$622 $\pm$ 66 & VLA \\ 21.96 & 4.86 &
$\phantom{0}$820 $\pm$ 52 & VLA \\ 15.90 & 1.43 & $\phantom{0}$250
$\pm$ 60 & VLA\\ 16.80 & 1.43 & $\phantom{0}$250 $\pm$ 70 & VLA\\
21.96 & 1.43 & $\phantom{00}$18 $\pm$ 56 & VLA \\ \enddata
\tablecomments{(1) Abbreviations: MAMBO = Max-Planck Millimeter
Bolometer; OVRO MA = Owens Valley Radio Observatory Millimeter Array;
PdBI = IRAM Plateau de Bure Interferometer; OVRO 40-m = Owens Valley
Radio Observatory 40-m telescope; Ryle = Ryle Telescope; VLA = Very
Large Array. (2) $\tau_{atm}$ = sky opacity toward the source.}
\end{deluxetable} 
\normalsize

\footnotesize
\begin{deluxetable}{llllll}
\tabcolsep0in\footnotesize \tablewidth{\hsize} \tablecaption{Results
of Spectral Fitting for \grb\,: the epoch of the spectral flux
distribution, the synchrotron self-absorption frequency $\nu_{\rm a}$,
the peak frequency $\nu_{\rm m}$, the peak flux density F$_{\rm m}$,
the reduced chisquared $\chi_{\rm r}^2$ and the degrees of
freedom (d.o.f.) of the fit.
\label{tab:fits}}
\tablehead { \colhead {Date} & \colhead {$\nu_{\rm a}$} & \colhead
{$\nu_{\rm m}$} & \colhead {F$_{\rm m}$} & \colhead {$\chi_{\rm r}^2$}
& \colhead {d.o.f} \\ \colhead {(UT)} & \colhead {(GHz)} & \colhead
{(GHz)} & \colhead {(mJy)} \\ } \startdata

1999 Dec 11.50 & \gpm{11.0}{1.5}{1.5} & \gpm{55}{28}{14} & $3.9 \pm
0.4$ & 0.0 & 2 \\

1999 Dec 13.50 & \gpm{7}{4}{4} & \gpm{35}{22}{12} & $3.9 \pm 0.5$ & 0.8 &
1 \\

1999 Dec 15.50 & \gpm{3.8}{1.2}{0.8} & \gpm{30}{12}{8} & $2.80 \pm
0.30$ & 1.0 & 3 \\

1999 Dec 21.50 & \gpm{8.8}{1.0}{1.6} & \gpm{4.8}{2.2}{1.4} & $2.10 \pm
0.25$ & 1.2 & 3 \\

\enddata \tablecomments{(1) In the spectral fitting
we have included the non detections at 1.4 and 250 GHz on 1999 Dec
21.5 UT as the peak flux density at the location of the afterglow. (2)
For Dec 21.5 UT we have also fitted using $\tau = 13/6$ (see
\S \ref{sec:specfits}) to account for the
fact that the peak frequency $\nu_{\rm m}$ is below the
self-absorption frequency $\nu_{\rm a}$ at this epoch and so F$_{\nu}
\propto \nu^{5/2}$. The result of the fit was the same within the
error bars.}
\end{deluxetable} 
\normalsize

\footnotesize
\begin{deluxetable}{cccccc}
\tabcolsep0in\footnotesize \tablewidth{\hsize} \tablecaption{Inferred
evolution of the synchrotron parameters $\nu_{\rm a}$, $\nu_{\rm m}$,
and F$_{\rm m}$ for \grb\, and the predicted scalings from the
constant density ISM,
wind and jet model. \label{tab:fitsres}} \tablehead { \colhead
{ } & \colhead {$\nu_{\rm a}$} & \colhead {$\nu_{\rm m}$} & \colhead
{F$_{\rm m}$} & } 

\startdata 

Observed & \gpm{12.0}{3.4}{2.7} $\times 10^9 \,t_{\rm obs}^{-0.15 \pm
0.12}$ Hz & \gpm{6.1}{7.1}{3.3} $\times 10^{11} \,t_{\rm obs}^{-1.7
\pm 0.4}$ \,Hz & \gpm{7.2}{1.6}{1.3} $\,t_{\rm obs}^{-0.47 \pm 0.11}$
\,mJy \\

& ($\chi_{\rm r}^2 = 5.5$; 2 d.o.f.) & ($\chi_{\rm r}^2 = 1.6$; 2
d.o.f.)  & ($\chi_{\rm r}^2 = 1.2$; 2 d.o.f.) \\

ISM & $\propto t_{\rm obs}^0$ & $\propto
t_{\rm obs}^{-3/2}$ & $\propto t_{\rm obs}^0$ \\

wind & $\propto t_{\rm obs}^{-3/5}$ & $\propto t_{\rm obs}^{-3/2}$ &
$\propto t_{\rm obs}^{-1/2}$ \\

jet & $\propto t_{\rm obs}^{-1/5}$ & $\propto t_{\rm obs}^{-2}$ &
$\propto t_{\rm obs}^{-1}$ \\

\enddata \tablecomments{(1) $t_{\rm obs}$ is the observer time in days
after the event. (2) The scalings for the jet model are valid if the
evolution is dominated by the sideways expansion of the jet (i.e. at
late times).
}
\end{deluxetable} 
\normalsize

\end{document}